\documentclass[journal]{IEEEtran}

\usepackage{microtype}                 
\PassOptionsToPackage{warn}{textcomp}  
\usepackage{textcomp}                  
\usepackage{mathptmx}                  
\usepackage{times}                     
\usepackage{cite}                      
\usepackage{tabu}                      
\usepackage{booktabs}                  
\usepackage[english]{babel}

\usepackage{amsmath,amssymb,amsfonts}
\usepackage{eucal}
\usepackage{graphicx}
\usepackage{xcolor}
\usepackage[linesnumbered,lined, boxruled,noend]{algorithm2e}
\usepackage{algorithmic}
\usepackage{setspace}
\usepackage{booktabs}
\usepackage{microtype}
\usepackage{hyperref}
\usepackage{bm}
\usepackage{listings}
\usepackage{color}
\usepackage[inline]{enumitem}
\usepackage[normalem]{ulem}
\usepackage{multirow}
\usepackage{subcaption}

\newcommand{\nonl}{\renewcommand{\nl}{\let\nl\oldnl}}

\newcommand{\sysname}{$\sf\small{FRAS}$\xspace}

\begin{document}

\title{FRAS: Federated Reinforcement Learning Empowered Adaptive Point Cloud Video Streaming}

\author{Yu Gao,
        Pengyuan Zhou,~\IEEEmembership{Member,~IEEE},
        Zhi Liu,~\IEEEmembership{Senior Member,~IEEE},
        Bo Han, ~\IEEEmembership{Senior Member,~IEEE},
        Pan Hui,~\IEEEmembership{Fellow,~IEEE}

\thanks{Yu Gao and Pengyuan Zhou are with the School of Cyber Science and Technology, University of Science and Technology of China, Hefei, China (e-mail: yugao@mail.ustc.edu.cn; pyzhou@ustc.edu.cn).}
\thanks{Zhi Liu is with Department of Computer and Network Engineering, The University of Electro-Communications, Tokyo, Japan (e-mail: liuzhi@uec.ac.jp).}
\thanks{Bo Han is with Department of Computer Science, George Mason University. (e-mail: bohan@gmu.edu) }
\thanks{Pan Hui is with Computational Media and Arts, Hong Kong University of Science and Technology, Guangzhou, China (e-mail: panhui@ust.hk).}
}

\maketitle
\begin{abstract}
Point cloud video transmission is challenging due to high encoding/decoding complexity, high video bitrate, and low latency requirement. Consequently, conventional adaptive streaming methodologies often find themselves unsatisfactory to meet the requirements in threefold: 1) current algorithms reuse existing quality of experience~(QoE) definitions while overlooking the unique features of point cloud video thus failing to provide optimal user experience, 2) most deep learning approaches require long-span data collections to learn sufficiently varied network conditions and result in long training periods and capacity occupation, 3) cloud training approaches pose privacy risks caused by leakage of user reported service usage and networking conditions.

To overcome the limitations, we present \sysname, the first federated reinforcement learning framework, to the best of our knowledge, for adaptive point cloud video streaming. We define a new QoE model which takes the unique features of point cloud video into account. Each client uses reinforcement learning~(RL) to train video quality selection with the objective of optimizing the user's QoE under multiple constraints. Then, a federated learning framework is integrated with the RL algorithm to enhance training performance with privacy preservation. Extensive simulations using real point cloud videos and network traces reveal the superiority of the proposed scheme over baseline schemes. We also implement a prototype that demonstrates the performance of \sysname via real-world tests.
\end{abstract}

\section{Introduction}
\label{sec:intro}

Volumetric video has become popular in recent years thanks to its immersive user experience with six degrees of freedom (6DoF), including the position (X, Y, Z) and the orientation (yaw, pitch, roll) of the viewer. Users can freely select any preferred viewing angle of the 3D scene to get an experience beyond 360-degree video restricted to 3DoF. In other words, unlike 360-degree video systems in which users can only switch orientation, users can freely move body position and head direction to subscribe to a field of view (FoV) at any location within the scene when watching a volumetric video.

Point cloud is arguably the most promising volumetric video format and has drawn significant attention from both academia and industry \cite{mekuria2016design,alexiou2017towards,schwarz2018emerging}. 
Delivering point cloud video requires network-friendly encoding, bandwidth-aware quality level adaption, efficient quality assessment metrics, and accurate 6DoF motion prediction~\cite{clemm2020toward,qian2019toward}. For instance, the required bandwidth for point cloud video streaming at 30 frames per second can be as high as 6 Gbps \cite{d20178i} due to the large size of raw point cloud frames, demanding efficient transmission methods.

Encoding and compression have thus been extensively researched to ensure transmission efficiency and user experience~\cite{schwarz2018emerging,cao20193d}. 
Tiling is popularly adopted as well, since a user can watch only a portion of the scene at a time~\cite{park2019rate,li2020joint}. For example, Park et al.~\cite{park2019rate} proposed partitioning the point cloud video into tiles spatially. Each tile is encoded with a different quality level according to its relation to the user's view frustum and distance to the user. 
%
%
However, decoding point cloud video requires more computation than traditional video~\cite{li2020joint,van2019towards,liu2021point}, posing a distinguished challenge for point cloud video systems. Several works consider using uncompressed tiles to alleviate the burden of en(de)coding. For example, Li et al.~\cite{li2020joint} considered the uncompressed tiles and optimized the user's quality of experience (QoE) by selecting the proper quality levels under the communication and computation constraints. Liu et al.~\cite{liu2020fuzzy} proposed a fuzzy logic solution to select the quality level for each tile based on the future bandwidth, user FoV, and the available computation capability. However, these schemes require predefined models or rules and thus cannot adapt to the dynamic network conditions. Although there are adaptive streaming approaches for traditional video streaming, such a proposal for point cloud streaming is yet to be explored.

\begin{figure*}[t!]
  \centering
    \begin{subfigure}[b]{0.4\textwidth}
       \includegraphics[width=\linewidth]{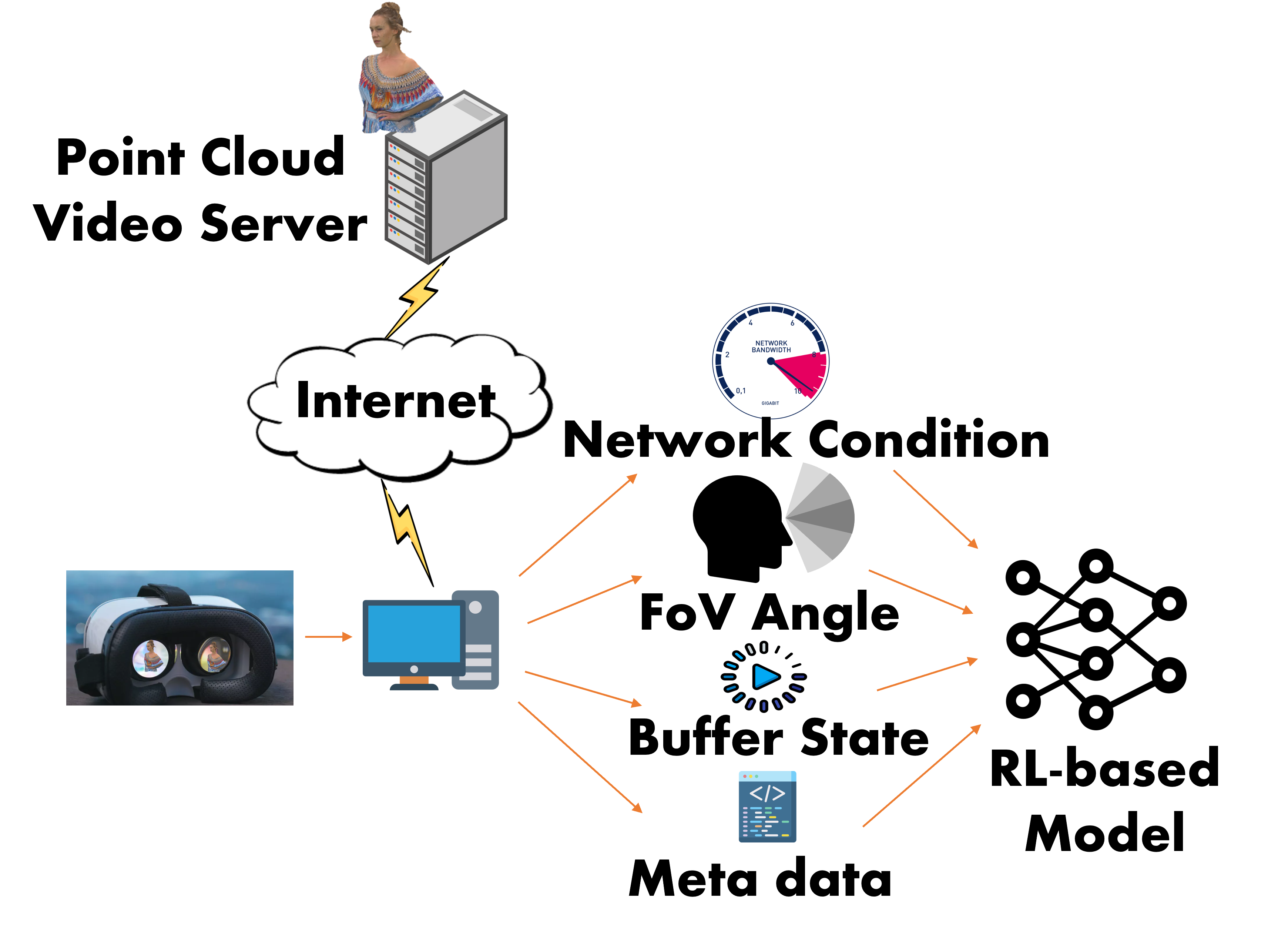}
       \caption{Local reinforcement learning.}
       \label{fig:teaser1}
    \end{subfigure}
    \begin{subfigure}[b]{0.4\textwidth}
       \includegraphics[width=\linewidth]{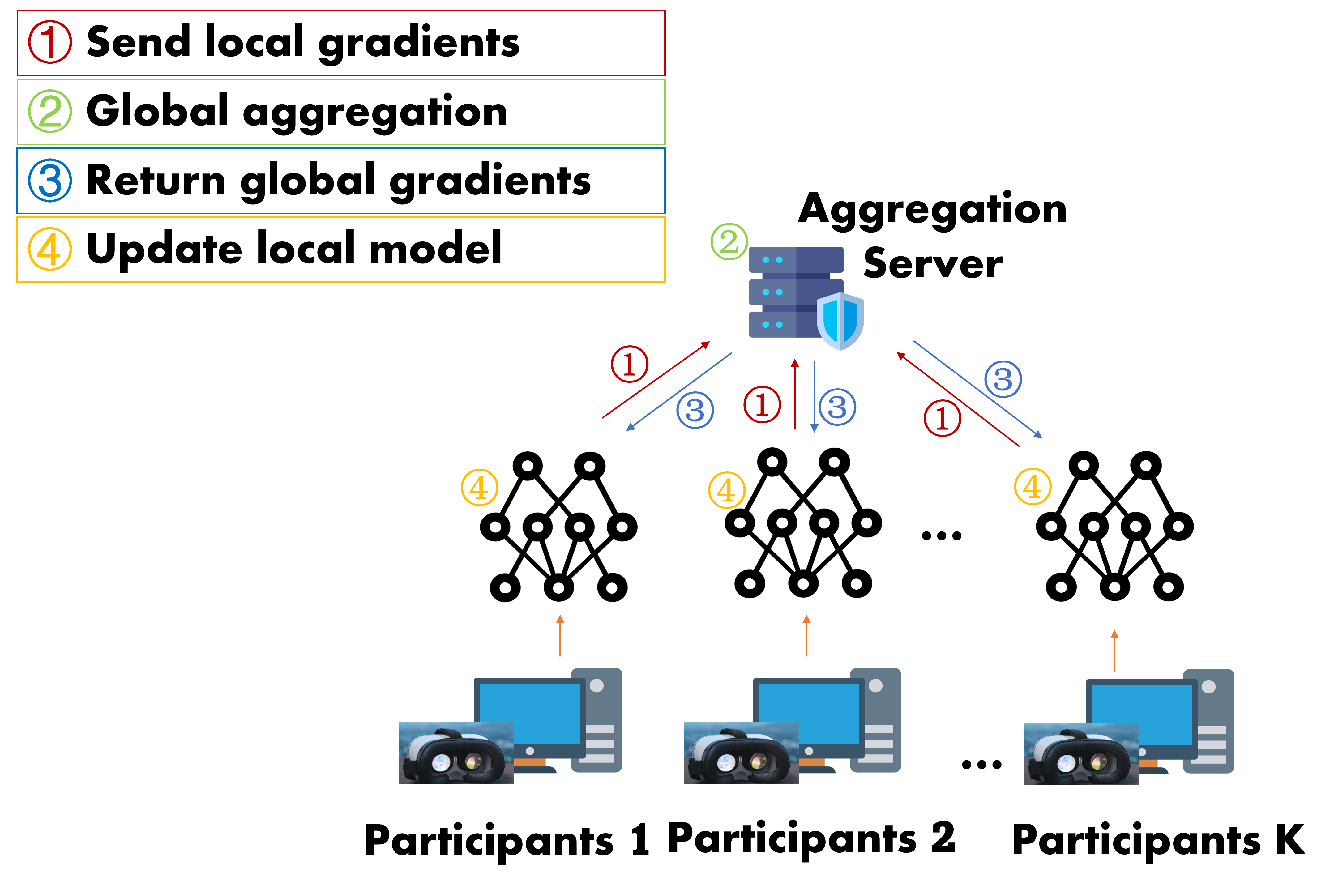}
       \caption{Global federated learning.}
       \label{fig:teaser2}
    \end{subfigure}
  \caption{An example to demonstrate the federal-reinforcement-learning-based adaptive point cloud streaming framework, i.e., \sysname: (a) Local training. Users choose the optimal video quality chosen by the RL model and update their local model. (b) Global training. Multiple users upload local gradients to a cloud server for federated model aggregation and retrieve the updated global model.}
  \label{fig:teaser}
\end{figure*}
To this end, in this paper we present \sysname, a federated reinforcement learning~(FRL) empowered point cloud video streaming system. The point cloud video is properly partitioned into tiles with different quality levels and different computation requirements for decoding. 
A novel quality evaluation metric for point cloud videos is proposed thereafter, which takes into account the features of a point cloud video system such as decoding complexity and viewer position. 
Then, based on the predicted network bandwidth and viewing direction and the decoding complexity, RL-empowered quality level selection helps maximize the perceived user experience under the constraints of communication resources, computation resources, and user quality requirements. To augment the training performance restricted by local datasets, \sysname applies federated learning to aggregate the learning experience of distributed clients while protecting their privacy. Extensive simulations based on real point cloud video sequences and network traces are conducted and the results reveal the superiority of the proposed scheme over baseline schemes including \textbf{ViVo}~\cite{han2020vivo},  \textbf{QUETRA}~\cite{yadav2017quetra}, \textbf{Pensieve}~\cite{mao2017neural}, \textbf{robustMPC}~\cite{yin2015control}, and \textbf{Buffer-Based} (BB)~\cite{huang2014buffer}. 
To the best of our knowledge, this is the first paper that investigates point cloud video streaming using FRL. Our contributions are summarized as follows:
\begin{itemize}[leftmargin=15pt]
\item We propose \sysname, a federated reinforcement learning~(FRL)-empowered adaptive point cloud streaming framework, as depicted in Figure~\ref{fig:teaser}. \sysname leverages distributed clients' datasets to improve the learning performance with privacy preservation. \sysname is the first FRL-based adaptive streaming (not only for point cloud video but in general) to the best of our knowledge. 
\item We propose a novel QoE definition that takes unique features of point cloud video into account. For example, we consider the decoding complexity of point cloud video, which is much higher than that of regular video, in the QoE model to make up for the rebuffering caused by decoding. Furthermore, we conducted a user study to validate our model and derive the weight parameters. 
\item Extensive simulations based on real point cloud video sequences and network traces are conducted, and the results reveal the superiority of the proposed scheme over baseline schemes. \textbf{\sysname outperforms all SOTA methods in all five concerned metrics}, specifically by up to 86\%, 28\%, 6\%, 11\%, and 67\% on the performances of average QoE, average quality level, average PSNR, average bandwidth, and rebuffering, respectively. We implement a prototype of \sysname and validate its performance via real-world tests.
\end{itemize}

\section{Related Work}
\label{sec:background}


\noindent \textbf{Point cloud encoding and processing.} A point cloud is composed of points represented in 3D space. Each point is associated with multiple attributes such as coordinates and color. There are two major classes of point cloud encoding methods according to point cloud data distribution~\cite{schwarz2018emerging,cao20193d}. A point cloud with uniform distribution can be projected into 2D frames using well-known 2D video technologies, while sparsely distributed point cloud data can be decomposed into hierarchical cubes with each point encoded as an index of its corresponding cube. Note that the point cloud encoding has higher computation complexity than the traditional video~\cite{qian2019toward,li2020joint}.


\vskip 0.1in\noindent \textbf{Point cloud quality assessment} defines the optimization objective and is an essential component of the point cloud video transmission system.
Peak signal-to-noise ratio (PSNR), which figures out the difference between the ground truth frame and the relieved frame pixel by pixel, is a common way for traditional video transmission to measure quality. For point cloud video quality assessment, researchers have proposed several metrics with similar logic. For example, \cite{hosseini2018dynamic} uses the PSNR of point-to-point distortions to measure the objective quality with the MPEG PCC\cite{mekuria2016overview} quality metric software. \cite{park2019rate} introduces utility measures based on the underlying quality of the representation, the level of detail to the user's viewpoint and device resolution. However, these schemes basically inherit the traditional video metrics and do not fully take the features of point cloud video into account.

\vskip 0.1in\noindent \textbf{Point cloud video streaming} has recently gained 
 popularity~\cite{li2020joint,clemm2020toward}. The majority of these works use similar methods to VR video streaming methods~\cite{zink2019scalable,guo2018optimal} that divide the videos into smaller tiles and only transmit the tiles within the user's FoV with the goal of optimizing the defined objective function~\cite{park2019rate,van2019towards}.
\cite{li2020joint}~considers the high computation complexity of point cloud video encoding during transmission optimization. These schemes are model-based and not adaptive to the dynamic network conditions.
\begin{figure}[t!]
    \centering
   \includegraphics[width=0.9\columnwidth]{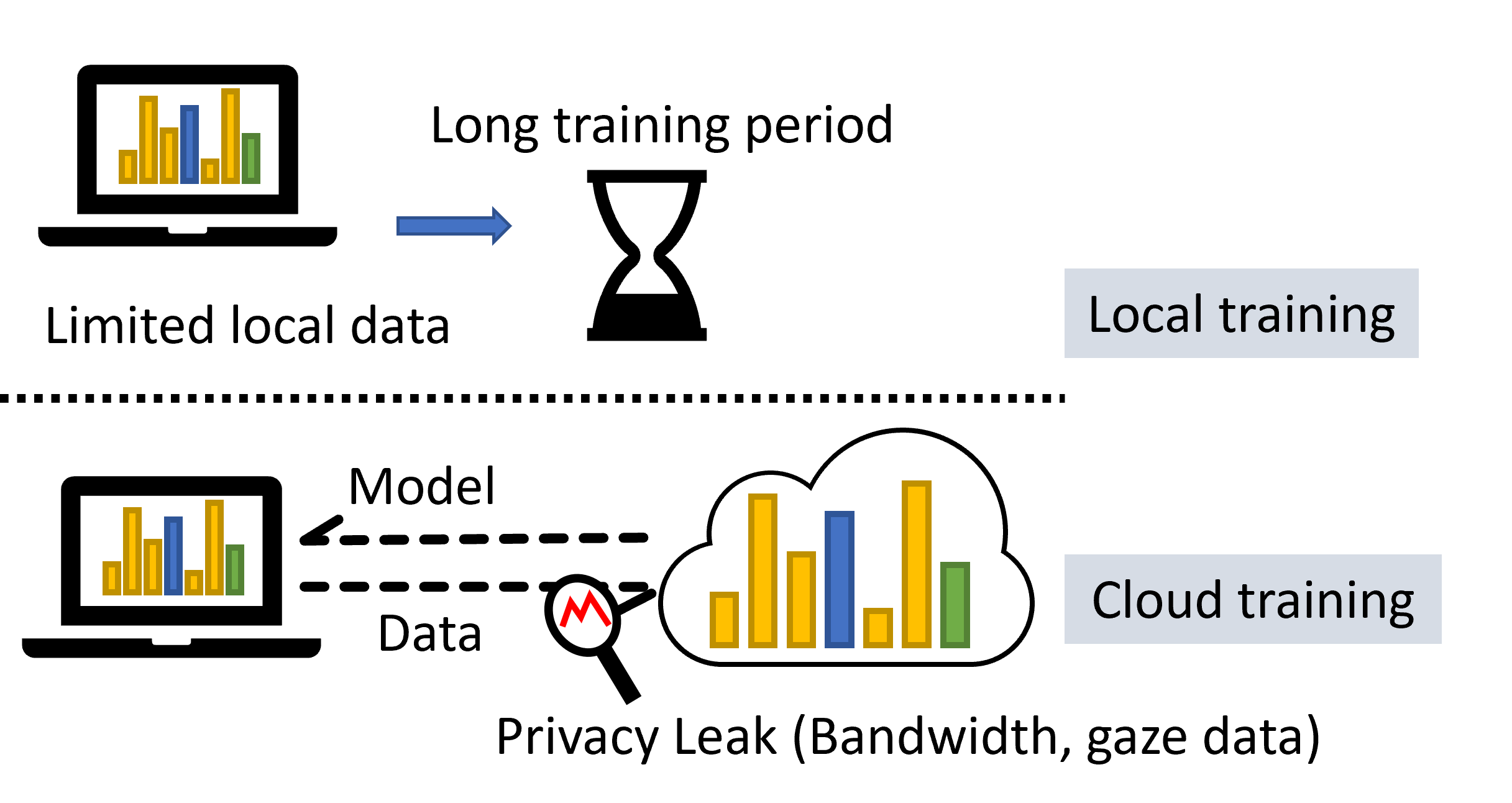}
    \caption{Shortcomings in existing approaches.}
   \label{fig:shortcoming}
\end{figure}

\vskip 0.1in\noindent \textbf{Adaptive bitrate streaming.} RL has become popular in traditional video streaming in recent few years~\cite{mao2017neural,xue2019new}. M. Claeys, et al.~\cite{claeys2014design} designed a RL-based HTTP adaptive streaming client interacting with the environment to optimize the QoE. The challenges in this direction include the proper definition of the reward and the training algorithm. Moreover, conventional local training provokes a significant limitation on the scale of the training dataset, specifically constrained to individual user devices. Since networking conditions change from session to session, each device needs a long time to collect data and learn about different networking conditions. As pointed out by Pensieve~\cite{mao2017neural}, retraining frequency depends on the generalizability of the model and the frequency of new network behaviors. The authors in Oboe~\cite{akhtar2018oboe} further demonstrate the necessity of retraining for Pensieve in the face of different network conditions. 

\vskip 0.1in\noindent \textbf{Privacy concern.} User privacy protecition is largely overlooked in previous works. The traffic metadata required for training, such as transmission rates, can seriously expose user privacy. For example, an attacker can infer user activity from traffic rate changes~\cite{apthorpe2017spying}. Furtheromre, a large amount of private information such as gender, age, cognitive state or even mental disorders can be revealed from user gaze data~\cite{sammaknejad2017gender,faber2018automated,hutton1984eye}.
Therefore, the conventional central cloud training methods that collect distributed user datasets put user privacy at risk.
\begin{figure*}[t!]
    \centering
   \includegraphics[width=\textwidth]{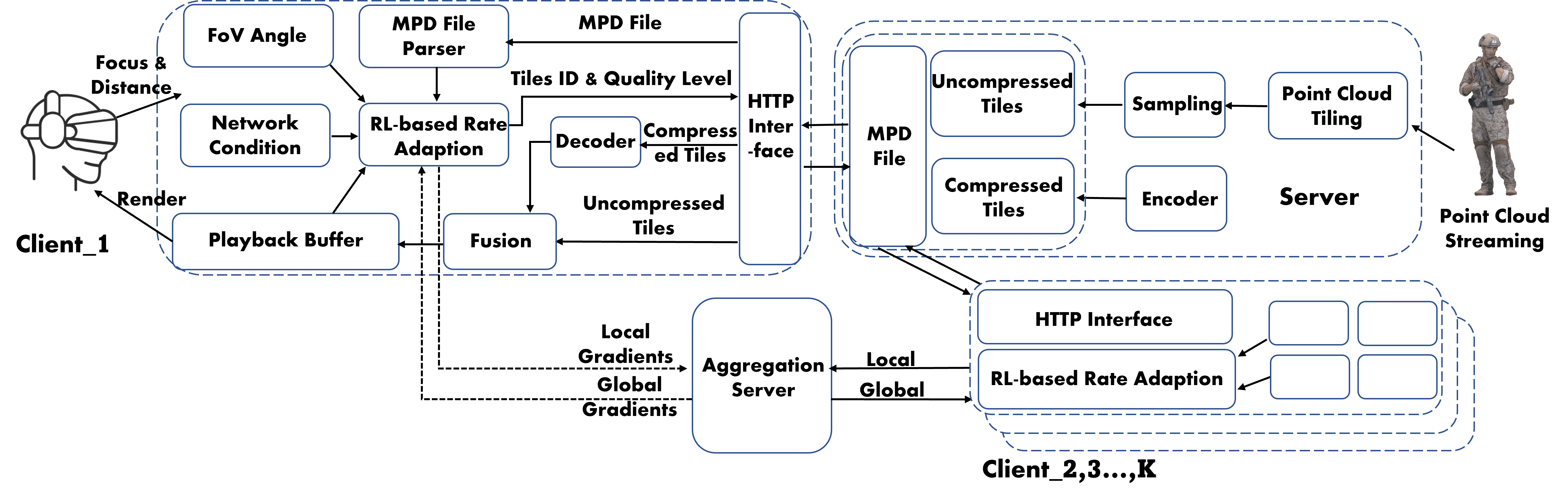}
    \caption{\sysname system architecture}
   \label{fig:system}
\end{figure*}

\vskip 0.1in To address the aforementioned shortcomings of related approaches as depicted in \autoref{fig:shortcoming}, we propose \sysname, a cross-device federated reinforcement learning adaptive streaming system to augment local model training via aggregating the learning experience from distributed clients. 
\sysname takes the decoding complexity and features of point cloud into the account of learning policy. Moreover, unlike current schemes relying on local user data~\cite{mao2017neural,yeo2018neural}, \sysname leverages uniquely diverse datasets from distributed users via federated learning to explore the generalizability of models extracted from global learning experience.
Without requiring user data uploading, \sysname guarantees privacy and non-identifiability for the resulting models~\cite{kairouz2019advances}. 

\section{\sysname System}
\label{sec:system}
This section introduces the tiling and quality assessment metric of point cloud videos.

\subsection{System overview}

\sysname is a video-on-demand system that adaptively delivers point cloud streaming, thus providing consumers with a greater quality of experience.
As illustrated in \autoref{fig:system}, 
the point cloud streaming is delivered to the server through a wired network and tiled. The partitioned tiles are sampled to different qualities and compressed. Then, the server uses all its tiles (both compressed and uncompressed) to generate an MPD file for each video and sends it to the client. Each client uses the RL-based rate adaption to select tiles ID and quality level based on environment information such as user's 6DoF pose, 
network condition, buffer state, and tile information obtained from the MPD file. Thereafter, the clients get the demanded tiles through the network interface. Meanwhile, multiple clients aggregate gradients via the aggregation server through the FRL algorithm during the training phase.

Because the bandwidth between video source and server is usually sufficient, we focus on the transmission between server and client which has limited and dynamic bandwidth $B_{i,t}$ for client $i$ at time point $t$. Note that $t$ is the basic time unit during which the frames are encoded together as a group, similarly to the group of pictures (GOP) in traditional video streaming. $B_{i,t}$ is predicted by employed bandwidth prediction methodology as explained in Section \ref{subsec:pre}. Consider client $i$ has a FoV $F_{i,t}$ at time point $t$ demanding video quality not lower than $Q_i$. User $i$ has $C_{i,t}$ available computation capacity to decode the compressed tiles. The client makes the decisions using RL based on the 6DoF pose, bandwidth, playback buffer, and the decoding process complexity according to the predefined reward policy to optimize QoE.

\subsection{Tiling and Downsampling}
\begin{figure}[t!]
    \centering
    \includegraphics[width=0.4\columnwidth]{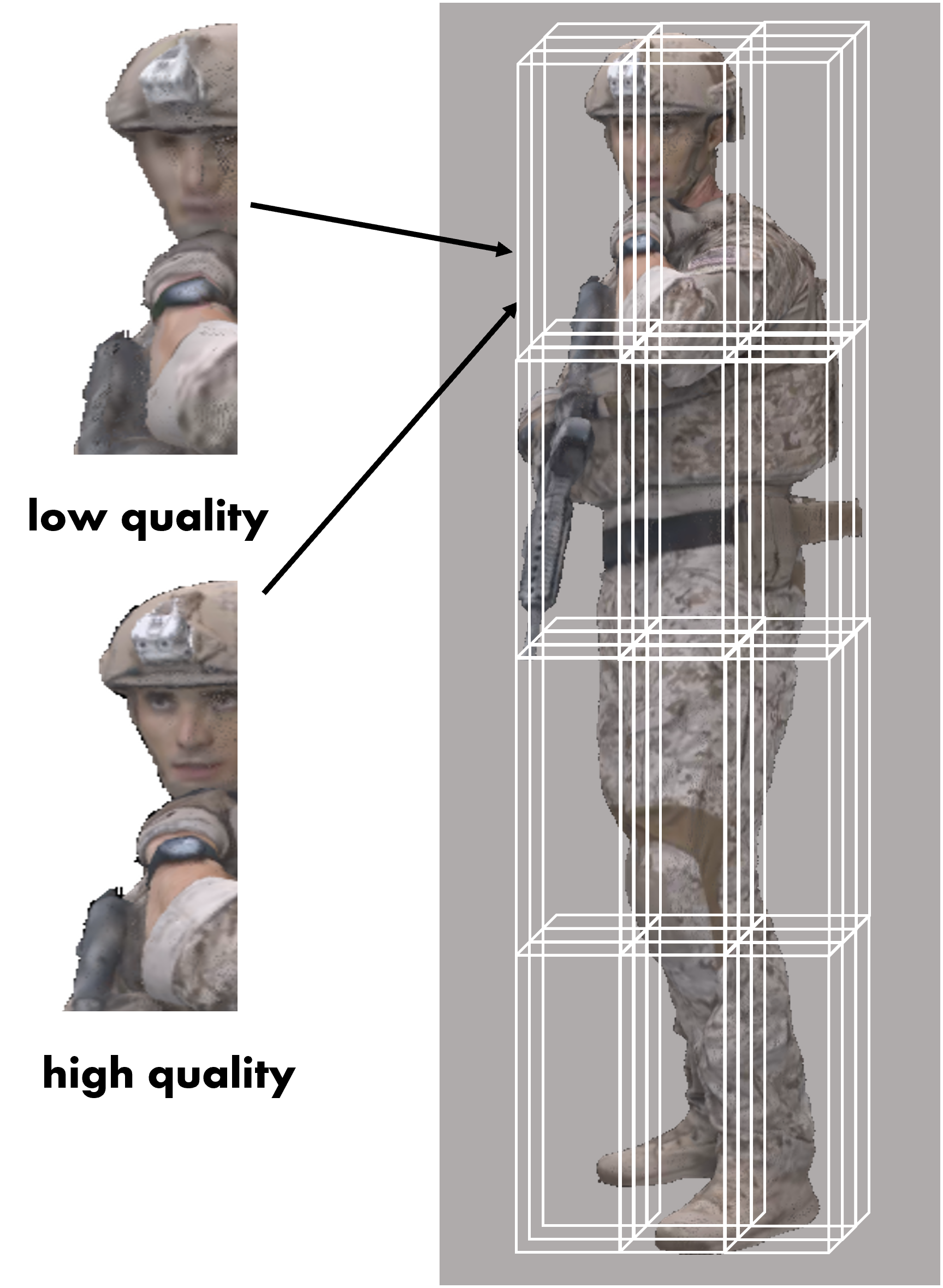}
    \caption{An example of tiling and two sampling qualities.}
    \label{fig:encoding}
\end{figure}

To guarantee the smooth viewport content switching and high video quality within the client's FoV, the point cloud video is partitioned into homogeneous tiles. For computation and communication efficiency, we transfer the tiles within FoV with the encoding rate selected by our FRL algorithm while encoding and transmitting the tiles outside FoV with the lowest quality.  
First, we recognize the best-fit cuboid surrounding the interested point cloud space and identify its pose and 3D dimension. Then, we perform $N \times M$ partitioning on the plane perpendicular to the height and divide them into $H$ layers in the height direction, and finally, $N \times M \times H$ tiles are obtained as shown in \autoref{fig:encoding}. The partitioned tiles at $t$ are further uniformly downsampled into $L$ quality levels. In this work, we downsample the tiles into 5 levels by 20\%, 40\%, 60\%, 80\%, and 100\%. A compressed tile $\{(n, m, h, l, t)| 1 \leq n \leq N, 1 \leq m \leq M, 1 \leq h \leq H, 1 \leq L\}$ has data size of $s_{n,m,h,l,t}$, point-to-point PSNR $q_{n,m,h,l,t}$, and required computation resource $d_{n,m,h,l,t}$ for decoding. A higher quality level has a larger data  size and requires more computation resources for encoding and decoding. 

We consider that the server also reserves uncompressed tiles at all quality levels. An uncompressed tile $(n, m, h, l, t)$ is with data size $s_{n,m,h,l,t}^{'}$, point-to-point PSNR $q_{n,m,h,l,t}$, and does not require computational capacity for decoding. We consider lossless compression in this work to avoid lossy compression's impact on depth data~\cite{wilson2017fast} and thus the uncompressed tiles have the same PSNR values as the compressed ones encoded at the same quality levels. The uncompressed tiles are at a larger size, i.e., $s_{n,m,h,l,t}^{'} > s_{n,m,h,l,t}$. Transmitting the uncompressed tiles allows for lower decoding workload on the client side at the cost of more bandwidth. 

\subsection{FoV and Network Prediction}
\label{subsec:pre}

While the $360^{\circ}$ video has only three degrees of freedom in terms of orientation (yaw, pitch, and roll), the point cloud video has three more degrees of freedom in terms of position (x, y, and z). This improves the user's immersive experience while increasing the complexity of FoV prediction. In this work, we use a simple and effective approach to predict the user's field of view, that is, to predict these six degrees of freedom separately and then combine them. We have tested three methods to predict each dimension: using last FoV, linear regression, and multilayer perceptron.

Network bandwidth prediction is also significant in point cloud video streaming.
In this work, we compare two methods for bandwidth prediction: gated recurrent unit~(GRU)~\cite{fu2016using} and exponentially weighted moving average~(EWMA)~\cite{hunter1986exponentially}. GRU is a popular artificial recurrent neural network architecture widely used for time-series forecasting.
It is similar to long short-term memory (LSTM) but with a forget gate, and can achieve slightly better performance with much lower complexity in comparison with LSTM. 
EWMA is a simple and powerful statistical tool for time series modeling.

\subsection{Constraints}
As mentioned, \sysname fetches all tiles at selective compression rates except those outside FoV which are transmitted without compression. 
We first define two notations to facilitate the description of the quality level selection. $I_{n,m,h,l,t}$ denotes the compressed tile $(n,m,h)$ at quality level $l$ during time period $t$, where $I \in \{0, 1\}$ represents the decision of transmission, $1$ transmitted and $0$ otherwise. Similarly, $I_{n,m,h,l,t}^{'}$ denotes the uncompressed tile $(n,m,h)$ at quality level $l$ during $t$.
We briefly list the constraints as follows. 

First, each transmitted tile is either compressed or uncompressed: 
\begin{equation}
\sum_{l=1}^{l=L}I_{n,m,h,l,t}+I_{n,m,h,l,t}^{'} = 1 , \forall n,m,h,t 
\label{constraint1}
\end{equation}
Second, at any time point $t$, the overall data volume of the transmitted tiles can not exceed the bandwidth:
\begin{equation}
\begin{aligned}
\sum_{n=1}^{n=N}\sum_{m=1}^{m=M}\sum_{h=1}^{h=H}\sum_{l=1}^{l=L}I_{n,m,h,l,t} \times s_{n,m,h,l,t} +I_{n,m,h,l,t}^{'} \times s_{n,m,h,l,t}^{'}  \leq B_{i,t}
\end{aligned}
\label{constraint2}
\end{equation}
Third, the required decoding resources for all transmitted tiles cannot exceed the capacity budget.
\begin{equation}
\sum_{n=1}^{n=N}\sum_{m=1}^{m=M}\sum_{h=1}^{h=H}\sum_{l=1}^{l=L}I_{n,m,h,l,t} \times d_{n,m,h,l,t} \leq C_{i,t}
\label{constraint3}
\end{equation}

\section{FRL-empowered transmission}
\label{sec:frl}

This section introduces the adopted federated reinforcement learning~(FRL) for adaptively selecting the proper tile quality level to maximize QoE under the communication and computation constraints. 

\subsection{Challenges}\label{subsec:challenge}
As the first look into this problem, we summarize the challenges as follows.
\begin{enumerate}[leftmargin=15pt]
    \item\label{challenge1} The decoding complexity is a unique feature of point cloud video compared to conventional video. For instance, decoding in point cloud video has longer latency than that in conventional video, and thus can affect the watching experience. We need to convert it to be comparable with other metrics to integrate it into the QoE calculation. 
    \item\label{challenge2} Another unique feature of point cloud video, 6DoF, enables clients to move positions. As a client would expect different frame qualities from different distances, the distance to the scene should also be considered. 
    \item\label{challenge3} Each round of federated learning consists of a number of local training epochs followed by the aggregation of clients' model updates. Therefore, a proper number of local training epochs is critical as too many would result in slow global model convergence while too few would lead to slow local model convergence and high communication cost.
    \item\label{challenge4} Lastly, though network prediction is not the focus of this work, it is an interesting additional functionality to improve ABR in general. Hence, it is important to ensure its processing delay does not impact the streaming service.
\end{enumerate}
\subsection{QoE Modeling Through User Study} \label{sec:qoe}
We mainly consider the point-to-point PSNR, encoding quality level, and distance from the scene for the quality metric. When a client moves closer to the scene, users will feel more sensitive to each tile's PSNR and its quality level. So the weights of quality components should vary with distance. Thus, for each client $i$ with FoV $F_{i,t}$ at time $t$, the quality of its received video is defined as follows:
\begin{equation}\label{eq:quality}
    Q_{dis} = \alpha * \sum_{FoV} q_{n,m,h,l,t} + \beta * l 
\end{equation}
which indicates the weighted sum of the encoding quality level and the sum of the point-to-point PSNR values of the tiles in the FoV, taking the user's distance into account. Next, the optimization goal also considers the rebuffering time $T_r$ and quality smoothness $\Delta l$ as in  related works. Moreover, as mentioned before, we take the decoding complexity of point cloud video into account. Since QoE already contains quality metrics and time metrics, adding a third-dimension metric would further increase its complexity and the training difficulty. Hence, instead of directly using the required decoding capacity, we include a decoding time penalty $T_d$ in the QoE, which refers to the actual decoding time subtracting the chunk duration time. We conducted numerous decoding tests across different point cloud videos of varied qualities and used the average values in the algorithm. \textbf{Notably}, the decoding time of point cloud video tile is much larger than that of conventional video which normally is considered an instantaneously part within $T_r$, as proved by our decoding tests. As such, the QoE of point cloud video streaming can be formulated as follows.

\begin{align} \label{eq:qoe}
     QoE_{dis}  & = f(Q,T_r,\Delta l, T_d) \\ \nonumber
     & = \alpha * \sum_{FoV} q_{n,m,h,l,t} + \beta * l - \gamma * T_r - \delta * \Delta l - \epsilon * T_d 
\end{align}
where $\alpha$, $\beta$, $\gamma$, $\delta$, and $\epsilon$ are weight parameters. 
Note that Our QoE model varies when the user's distance to the FoV changes. 
Therefore, we derive groups of weight values according to different distances, as explained later, according to our user study results. The problem is to maximize the average QoE. As such, we solve \textbf{challenge \ref{challenge1} and challenge \ref{challenge2}}. 

Next, we conduct a user study to validate our defined QoE model and derive the weight parameters. Our approach to studying the impact of all factors in the QoE model mentioned in Equation~\eqref{eq:qoe} is to have each participant compare the videos impaired by various combinations of factors with a high-quality ``perfect'' version  ($\overline{Q}=5$, $\overline{T}_r=\overline{\Delta l}=0$) in each distance and provide a subjective rating. We choose the \textit{Longdress} from the 8i Labs point cloud datasets\cite{d20178i} for this study, which shows a dancing woman and lasts for 10 seconds. \autoref{tab:factorsl} shows the factors and their possible values for our QoE model training. There are 108 combinations of video pairs, with the impaired version on the left and the perfect version on the right. Then we ask each participant to randomly select 36 video pairs from the 108 combinations for a subjective experience comparison. After watching each pair, the participant rates which QoE the impaired video provides compared to the perfect video through 4 choices: similar to, slightly worse, worse, much worse. Then the impaired video's QoE is labeled with a score of (3,2,1,0) respectively.
\begin{table}[htbp]
    \renewcommand\arraystretch{1.2}
    \centering
    \setlength{\tabcolsep}{4.5mm}
    \begin{tabular}{ll}
        \toprule
        factor & possible values \\
        \midrule
        Avg. video quality $\overline{Q}$ & {1, 2, 3, 4, 5}\\
        Avg. distance $\overline{dis}$ & {1m, 2m, 3m}\\
        Avg. rebuffer time $\overline{T}_r$ & {0.00s, 0.25s, 0.50s, 1.00s}\\
        Avg. quality smoothness $\overline{\Delta l}$ & {0, 1, 2}\\
        \bottomrule
    \end{tabular}
    \caption{The factors and their possible values for QoE model modeling.}
    \label{tab:factorsl}
\end{table}

\begin{figure}[t!]
    \centering
       \includegraphics[width=.6\linewidth]{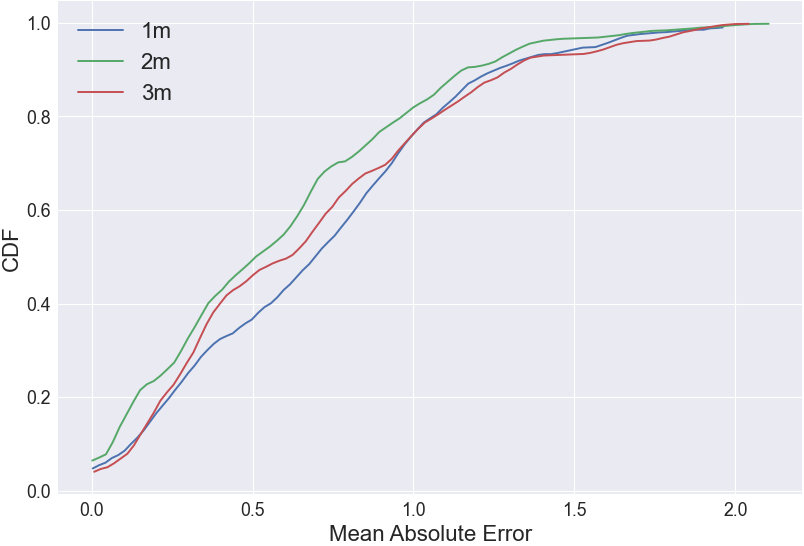}
    \caption{Mean absolute error in QoE model validation.}
    \label{fig:user_study}
    \vspace{-0.1 in}
\end{figure}

We collected 1080 ratings from 30 participants through a survey and then used the ratings to calculate the QoE parameters in Equation~\eqref{eq:qoe} for each distance (1\,m, 2\,m, 3\,m) separately through linear regression. The results are shown in \autoref{tab:qoe_params}. Then we used 10-fold cross-validation to validate our QoE model and the mean absolute error (MAE) of the prediction results is shown in~\autoref{fig:user_study}. The average MAE at 1m, 2m, 3m is 0.71, 0.59, and 0.68, respectively.
\begin{table}[htbp]
    \renewcommand\arraystretch{1.2}
    \centering
    \setlength{\tabcolsep}{3mm}

    \begin{tabular}{|c|c|c|c|c|c|}
        \hline
        Distance& $\alpha$ & $\beta$ & $\gamma$ & $\delta$ & $\epsilon$ \\
        \hline
        1\,m & 0.11 & 0.61 & 12.58 & -0.13 & 12.58  \\
        2\,m & 0.05 & 0.12 & 12.68 & -0.01 & 12.68 \\
        3\,m & 0.04 & 0.10 & 13.29 & -0.05 & 13.29 \\
        \hline
    \end{tabular}
    \caption{QoE model weight values used in \sysname.}
    \label{tab:qoe_params}
    \vspace{-0.1 in}
\end{table}

\subsection{FRL Algorithms}
We train local models using RL algorithm to optimize each individual client's QoE following the objective function defined in \autoref{sec:qoe}. As mentioned, each client encounters limited network conditions in short periods and thus needs long-span data collection and training to get a generalized and robust model, which impacts the QoE of the client and the battery life of the device. Therefore, we integrate FL and RL to aggregate the learning experiences of users for data augmentation and training acceleration under privacy preservation. The major details are listed as follows. 

\vskip 0.06in\noindent \textbf{State}:
The RL agent takes $s_t = (l_{t-1}, b_t, B_t, d_t, n_t, r_t)$ as the input, where $l_{t-1}$ is the selected encoding quality level of the last point cloud video chunk 
(uncompressed tiles have different level indices than compressed tiles to differentiate); $b_t$ is the current buffer level; $B_t$ is the predicted network bandwidth; $d_t$ is the download time of the current chunk; $n_t$ is a vector of $L$ possible sizes for the next video chunk; $r_t$ is the number of remained chunks in the video.

\vskip 0.06in\noindent\textbf{Action}: The RL agent takes an action by selecting the encoding quality level for the next point cloud video chunk upon receiving $s_t$. The actions are selected based on a policy $\pi_\theta : \pi_\theta (s_t, a_t) \rightarrow [0, 1] $

\vskip 0.06in\noindent\textbf{Local training}:  We employ actor-critic algorithm to train the local policy. The reward is defined according to QoE~(\autoref{sec:qoe}) as shown in Equation~\eqref{eq:reward}. The gradient of the cumulative discounted reward with respect to the policy is computed as Equation~\eqref{eq:gradient}~\cite{mnih2016asynchronous}.
\begin{equation}\label{eq:reward}
\begin{aligned}
    r_t= QoE
\end{aligned}
\end{equation}
\begin{equation}\label{eq:gradient}    
    \nabla_{\theta} \mathbb{E}_{\pi_{\theta}}\left[\sum_{t=0}^{\infty} \gamma^{t} r_{t}\right]=\mathbb{E}_{\pi_{\theta}}\left[\nabla_{\theta} \log \pi_{\theta}(s, a) A^{\pi_{\theta}}(s, a)\right]
\end{equation}

\vskip 0.06in\noindent\textbf{Global aggregation}: We apply FedAvg\cite{mcmahan2017communication} in the global aggregation stage. After $E$ local epochs, $m$ of $K$ clients return their local gradients to the aggregation server, where $m$ equals to $max(\mu *K,1)$ and $\mu$ is a pre-defined participation ratio. Then the global gradients are calculated according to Equation~\eqref{eq:global} and used by all the K clients to update local models. 
\begin{equation}\label{eq:global}
    \theta_t^G = \sum_{k=1}^{\mu * K}\frac{w_k*\theta_t^k}{\mu * K}
\end{equation}
where $\theta_t^G$ and $\theta_t^k$ indicate the global and local gradients, $w_k$ is the weight of client k.

\IncMargin{0.5em}
\begin{algorithm}[!t]
 \SetKwProg{proc}{Procedure}{:}{}
 \SetKwProg{func}{function}{:}{}
   \proc{FL}{
    \KwIn{Server actor network weight $\theta_0^G$ and critic network weight  $\theta_{v_0}^G$, users $\langle c_1,c_2,...c_K \rangle$, local update step $E$}
      \For{$T=1$ to $T_{max}$}{
        Select a number of users $\mathbf{C} = \langle C_1^T,...C_{CK}^T \rangle$\;
        $u_{T+1}^\mathbf{C},u_{v_{T+1}}^\mathbf{C} \gets \textit{RL}(\theta_{T}^G,\theta_{v_{T}}^G) $ \;
        $u_{T+1}^G \gets \textit{FedAvg}(u_{T+1}^\mathbf{C})$\;
        $u_{v_{T+1}}^G \gets \textit{FedAvg}(u_{v_{T+1}}^\mathbf{C})$\;
       Update global model $\theta_{T+1}^G \leftarrow \theta_{T}^G+u_{T+1}^G$, $\theta_{v_{T+1}}^G \leftarrow \theta_{v_{T}}^G+u_{v_{T+1}}^G$\;
      }
     }
  \proc{RL}{
      \KwIn{$ \mathbf{C} = \langle C_1^t,...,C_{CK}^t \rangle, \text{ global model } \theta_{T}^G, \theta_{v_T}^G $}
        \For{each user $c \in \mathbf{C}$}{
              Reset $d\theta$ and $d\theta_v$ to 0, $\theta \gets \theta_T^G$, $\theta_v \gets \theta_{v_T}^G$ \;
              $\theta' = \theta$, $\theta_v' = \theta_v$\;
              \For{$t = t_{start}$ to $t_{start+E-1}$}{
              $FoV_t$, $B_t$ = Prediction($\mathbf{c}$)\;
              $n_t \gets f(FoV_t)$ \;
              $s_t = (l_{t-1}, b_t, B_t, d_t, n_t, r_t)$ \;
              Perform $a_t$ according to $\pi (a_t|s_t;\theta')$\;
              Receive reward $r_t$ and new state $s_{t+1}$
              }
              \For{$t \in \{t_{start+E-1},...,t_{start}\}$}{ 
             $d\theta \gets d\theta + \nabla_{\theta '}log(\pi (a_{i}|s_{i};\theta ')(R-V(s_{i};\theta_{v}')))$\;
             $d\theta _{v} \gets d\theta _{v} + \frac{\partial ((R-V(s_{i};\theta _{v}'))^{2})}{\partial \theta _{v}'}$\;}
             $t_{start} \gets t_{start+E}$\;
             Update $\theta$ and $\theta_v$\;
             Return $d\theta$ and $d\theta_v$
            }
    }
    \proc{Prediction}{ 
              $FoV \gets FoVpredict(c)$ \;
              $B \gets BWpredict(c)$ \;
              Return $\mathbf{c}_{FoV}$, $\mathbf{c}_{B}$\;
         }    

\caption{\sysname}
\label{alg1}
\end{algorithm}
\DecMargin{0.5em}

Algorithm~\ref{alg1} summarizes the overall functionalities. In detail, before each round, the server selects a ratio of the clients to participate in the training. Each selected client trains a local actor-critic model using its own dataset. Before each training step, the client predicts the FoV and bandwidth using the methods described in \autoref{subsec:pre}. The client uses the predicted FoV and possible encoding quality levels to calculate the possible sizes of the next video chunk. 
As such, each client collects the current state $s$, performs an action according to the policy $\pi$, and accumulates the reward. Using the forward view and mix of n-step returns methods of the actor-critic algorithm, each client accumulates the gradients of its policy and value-function. Every $E$ epochs, the policy and value-function are updated and the policy gradients are uploaded to the server for aggregation. Upon receiving sufficient updates, the server conducts FedAvg over the gradients and updates the global model. Then the server sends the global model to another batch of selected clients and starts the next round of training until the global model converges.

Note that the integration of FL and RL is not straightforward and requires fine-tuning efforts. As stated in \textbf{challenge \ref{challenge3}}, the $E$ local epochs of RL training affect the trade-off between transmission cost and convergence rate of local and global models. We found the optimal $E$ for different data distributions via empirical tests. For \textbf{challenge \ref{challenge4}}, to enable real-time network prediction, we customize the input/output period lengths to decrease the processing delay to a level that can be transparent to the user while still facilitating RL training and inference. Please refer to Section \ref{sec:evaluation} for the details. Another ad-hoc configuration is the number of workers ran by each client's RL algorithm. Depending on the available capacity, each RL algorithm can run a varied number of workers in parallel using an asynchronous actor-critic algorithm~(A3C~\cite{mnih2016asynchronous}), or, just one single worker using synchronous update~(A2C~\cite{wu2017scalable}). The performance of both methods is comparable, as found by the empirical study~\cite{wu2017scalable}. A few newer RL algorithms can also be applied in this context, which is not our focus and thus left out of this work.


\section{Evaluation}
\label{sec:evaluation}

\subsection{Experimental Setup}
\noindent \textbf{Video source.} We use 36 point cloud video sequences from Panoptic Studio\footnote{\url{http://domedb.perception.cs.cmu.edu/ptclouddb.html}} for training. We use another 6 point cloud video sequences from vsenseVVDB2 database \footnote{\url{https://v-sense.scss.tcd.ie/research/6DoF/quality-assessment-for-fvv-compression/}}, i.e., \textit{AxeGuy, LubnaFriends, Rafa2, Matis, Loot} and \textit{ Longdress}, for testing. Each point cloud frame is partitioned into $3 \times 3 \times 4$ tiles and then sampled at (20\%, 40\%, 60\%, 80\%, 100\%), corresponding to quality level from 1 to 5. We use the KD-tree based Google Draco codec\footnote{\url{https://github.com/google/draco}} for encoding and decoding. We use the Draco rather than standardized MPEG's V-PCC encoder\footnote{\url{https://github.com/MPEGGroup/mpeg-pcc-tmc2}} because it has better performance and can allow real-time decoding on contemporary hardware. 


\vskip 0.08in \noindent \textbf{Network conditions.} Since the point cloud video has high bitrates even after compression, we must use high-bandwidth traces to emulate realistic throughput fluctuations in the experiments. Thus, we replay the bandwidth traces from a commercial mmWave 5G throughput dataset provided by Lumos5G~\cite{10.1145/3419394.3423629} to simulate the network. The dataset contains 118 traces under two different mobility modes: walking and driving, covering over 18 hours of active monitoring.  During the test, the algorithm uses predicted bandwidth as described in \autoref{subsec:pre} to select the encoding quality for the next video chunk. Then the algorithm uses the selected quality level and predicted bandwidth together with several other state metrics to calculate the reward.

\vskip 0.08in\noindent \textbf{FoV.} In this work, we use the viewport trajectory datasets provided by ViVo~\cite{han2020vivo}, which provide the viewport trajectory of 32 users (including smartphone users and headset users) in four volumetric videos. The datasets have position dimensions (x, y, z) and orientation dimensions (yaw, pitch, roll) in each item and have over 10,000 items overall. As mentioned in \autoref{subsec:pre}, we use three approaches to separately predict each dimension: 1) last FoV (LAST): use the FoV of the previous moment as the current FoV; 2,3) LR, MLP: use LR or MLP to predict the current FoV from a history window consisting of last 8 viewports. The lightweight neural network we use in MLP has two hidden layers with 8 and 2 neurons, respectively. The activation function of MLP is rectified linear unit (ReLU), and the solver for weight optimization is Limited-memory BFGS (L-BFGS)~\cite{liu1989limited}. Additionally, LR and MLP are both implemented by the scikit-learn library\footnote{https://scikit-learn.org/}.

\vskip 0.08in\noindent \textbf{Models and Metrics.} \sysname passes 12 past bandwidth measurements to a 1D convolution layer with 128 filters, each of size 4 with stride 1. Possible next chunk sizes are input to another 1D-CNN with the same shape. Results from these layers are then aggregated in a 128-neuron hidden layer to apply the softmax function. The critic network uses the same structure but a linear neuron as the final output. We set the discount factor $\gamma=0.99$, the learning rates at $10^{-4}$ and $10^{-3}$ for the actor and critic, and, the entropy factor $\beta$ decay from 5 to 0.1 over $3*10^5$ iterations. We implemented this architecture using TensorFlow 2.3.0. Please refer to \autoref{tab:qoe_params} for the QoE model weight values we used.

\vskip 0.08in\noindent \textbf{Baselines.} We implement 5 baselines for performance comparison. (1) ViVO\cite{han2020vivo} performs a first comprehensive study of mobile volumetric video streaming, which considers the 6DoF feature of volumetric video and optimize the streaming from three visibility-aware approaches. (2) QUETRA~\cite{yadav2017quetra} is a DASH rate adaptation algorithm which calculates the expected buffer occupancy using selected bitrate, network throughput, and buffer capacity based on queuing model. (3) An RL method adapted from Pensieve~\cite{mao2017neural} which trains a single RL model. (4) robustMPC~\cite{yin2015control} improves MPC by accounting for errors of predicted throughput via normalization. (5) Buffer-Based (BB)~\cite{huang2014buffer} selects the encoding quality level based on the current buffer occupancy and estimated occupancy during the startup phase with the goal of keeping the occupancy above 0.1 second, and automatically chooses the highest level if the occupancy exceeds 1 second.

\vskip 0.05in\noindent \textbf{Implementation.} We built a point cloud visualizer using PCL 1.9.1 and QT 5.12 to support each of the aforementioned baselines. The visualizer was configured to fetch quality selection decisions from an ABR client that implemented the corresponding algorithm. The player was configured to have a playback buffer capacity of 5000\,ms. Each sequence has 300 frames, or 10 seconds of video, and will be played in a loop to present a more realistic playback scenario. Furthermore, we set the GOP to 10 in experiments, i.e., each chunk has approximately 330\,ms duration in order to avoid sending a large number of requests in a very short time. 
The video player and the ABR server run on the same server as the client, which is equipped with an Intel Xeon Gold 6246R CPU, NVIDIA GeForce GTX 3090 GPU and 480G SSD. And an HTTP server was deployed with Python HTTP module, which stored tiled point cloud videos and corresponding MPD files. 

\subsection{Result and Analysis}

\begin{figure}[t!]
    \centering
    \begin{subfigure}[b]{0.6\columnwidth}
       \includegraphics[width=\linewidth]{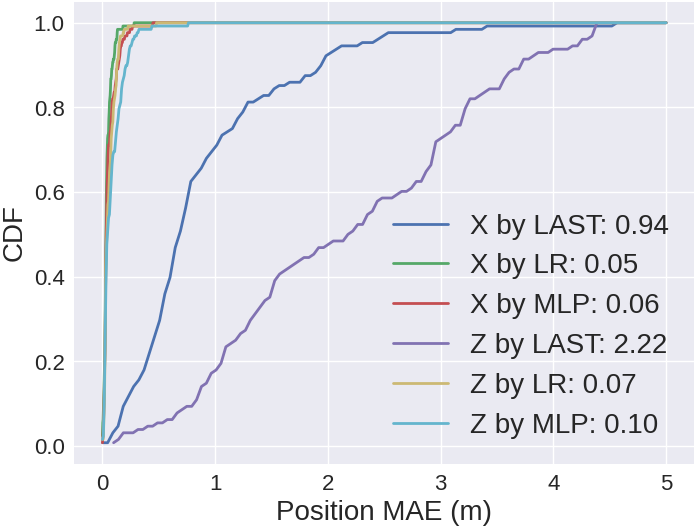}
       \caption{Position prediction.}
       \label{fig:fovp}
    \end{subfigure}
    \begin{subfigure}[b]{0.6\columnwidth}
       \includegraphics[width=\linewidth]{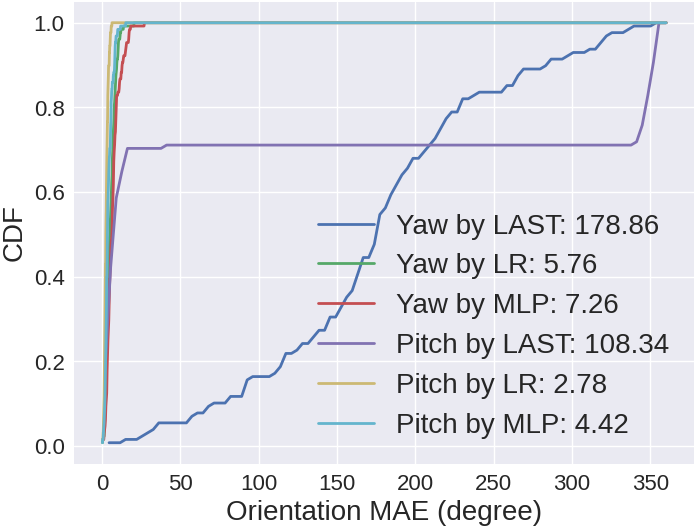}
       \caption{Orientation prediction.}
       \label{fig:fovo}
    \end{subfigure}
    \caption{FoV prediction.}
    \label{fig:fov}
\end{figure}

\begin{figure}[t!]
    \centering
    \begin{subfigure}[b]{0.6\columnwidth}
       \includegraphics[width=\linewidth]{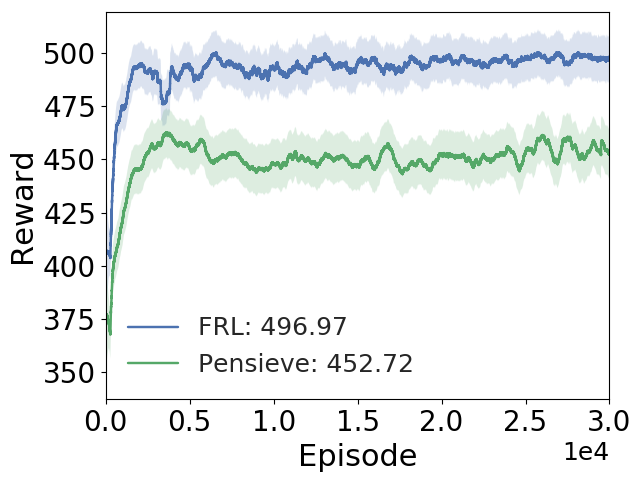}
       \caption{Reward}
       \label{fig:training1}
    \end{subfigure}
    \begin{subfigure}[b]{0.6\columnwidth}
       \includegraphics[width=\linewidth]{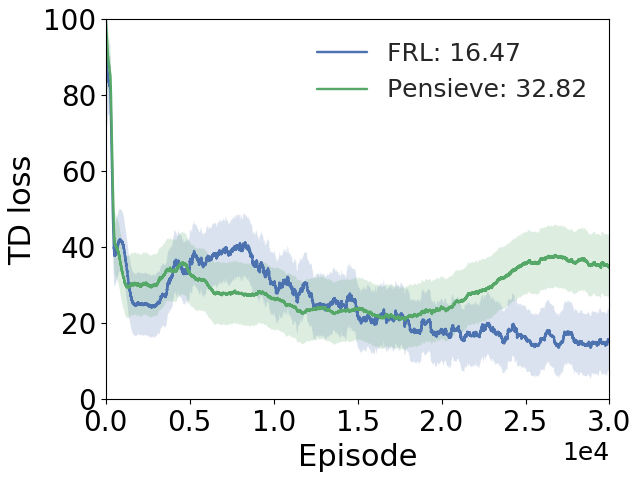}
       \caption{Loss}
       \label{fig:training2}
    \end{subfigure}
    \caption{Comparison of FRL and Pensieve in training.}
    \label{fig:training}
    \vspace{-0.1 in}
\end{figure}
\begin{figure}[t!]
    \centering
       \includegraphics[width=\linewidth]{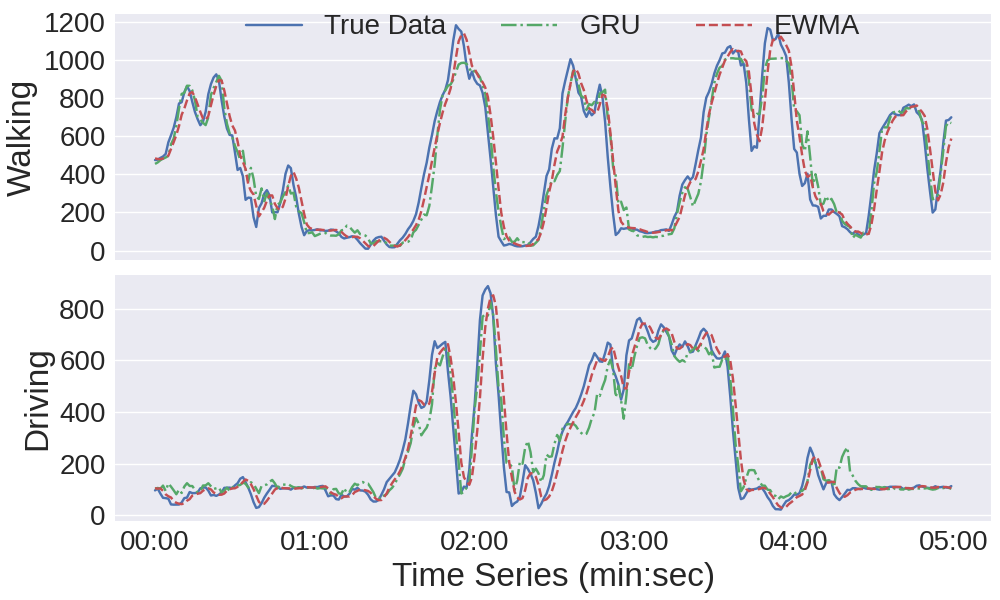}
    \caption{Network prediction performances of GRU and EWMA in different networking conditions: walking and driving. Y axis indicates bandwidth in Mbps.}
    \label{fig:netprediction}
    \vspace{-0.2 in}
\end{figure}
\begin{figure}[t!]
    \centering
       \includegraphics[width=\linewidth]{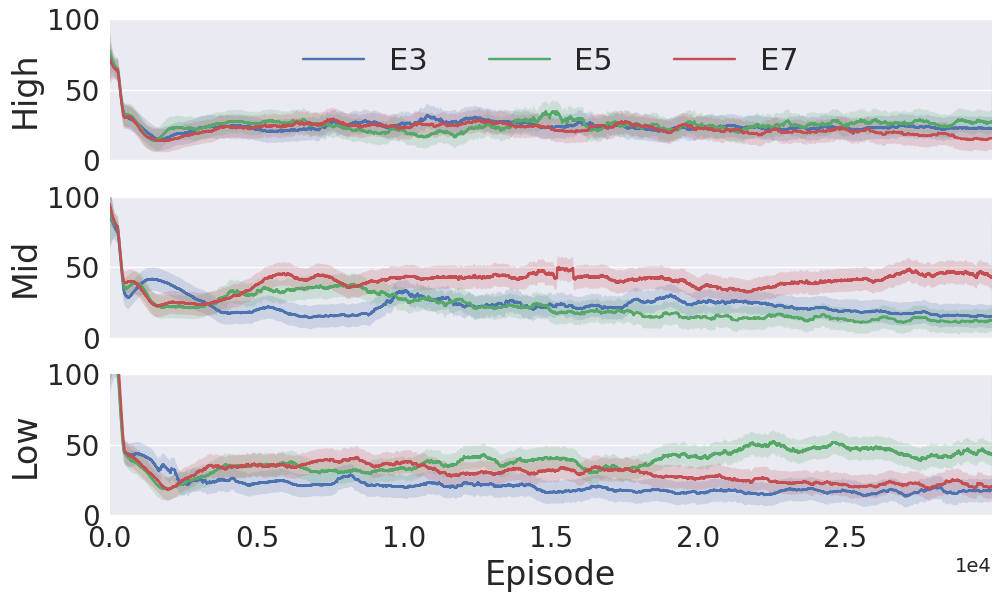}
    \caption{Affect of $E$ in different data distribution heterogeneities. Y axis indicates loss.}
    \label{fig:hetero}
\end{figure}
\begin{figure*}[t!]
    \centering
    \begin{subfigure}[b]{0.32\textwidth}
       \includegraphics[width=\linewidth]{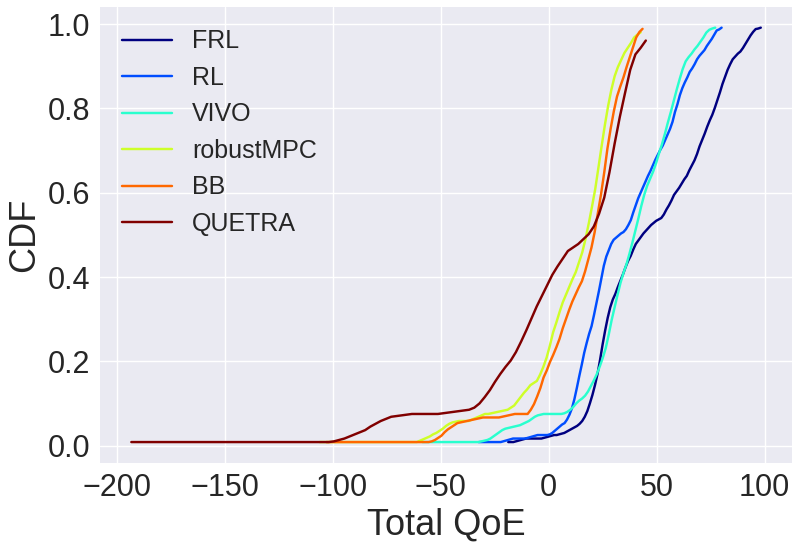}
       \caption{Sum of QoE}
    \end{subfigure}
    \begin{subfigure}[b]{0.32\textwidth}
   \includegraphics[width=\linewidth]{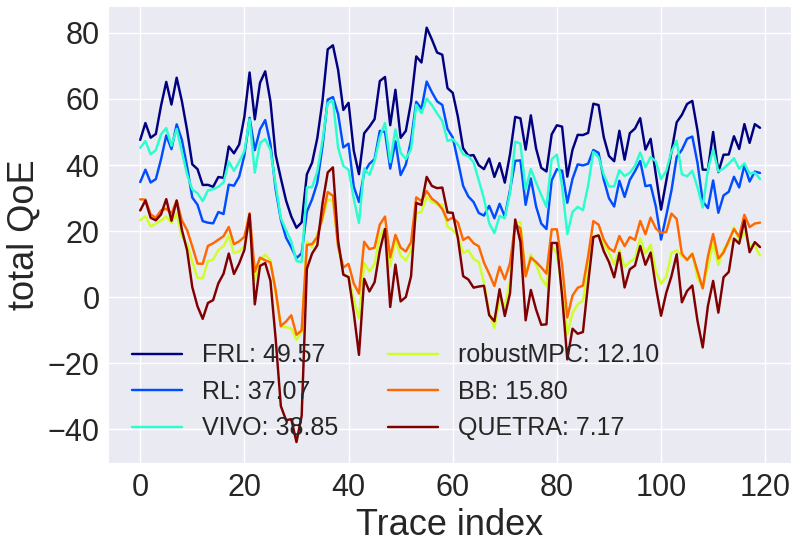}
   \caption{Average QoE}
    \end{subfigure}
    \begin{subfigure}[b]{0.32\textwidth}
       \includegraphics[width=\linewidth]{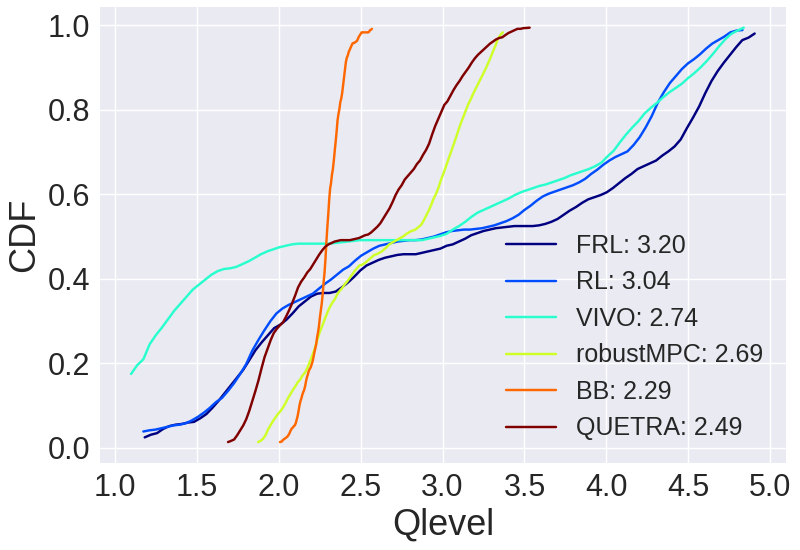}
       \caption{Sum of quality level}
    \end{subfigure}
    \caption{QoE performance.}
    \label{fig:testperform}
\end{figure*}

\begin{figure*}[t!]
    \centering
    \begin{subfigure}[h!]{0.32\textwidth}
       \includegraphics[width=\linewidth]{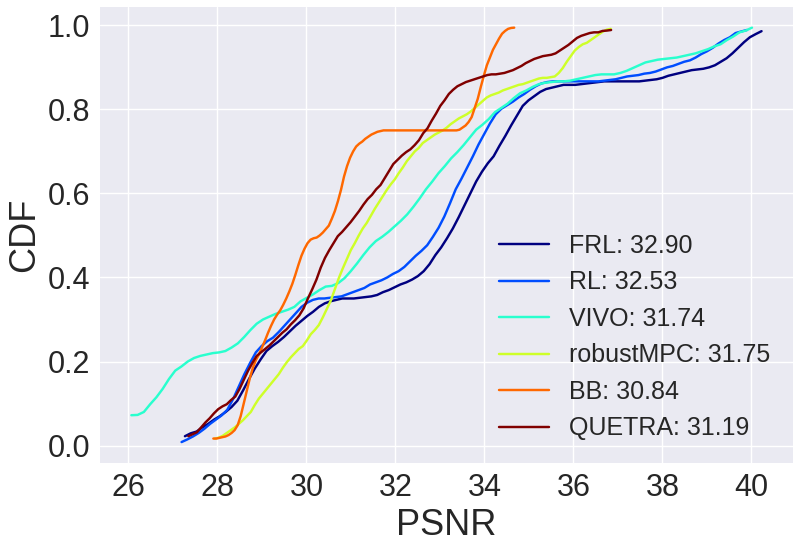}
       \caption{PSNR}
    \end{subfigure} 
    \begin{subfigure}[h!]{0.32\textwidth}
       \includegraphics[width=\linewidth]{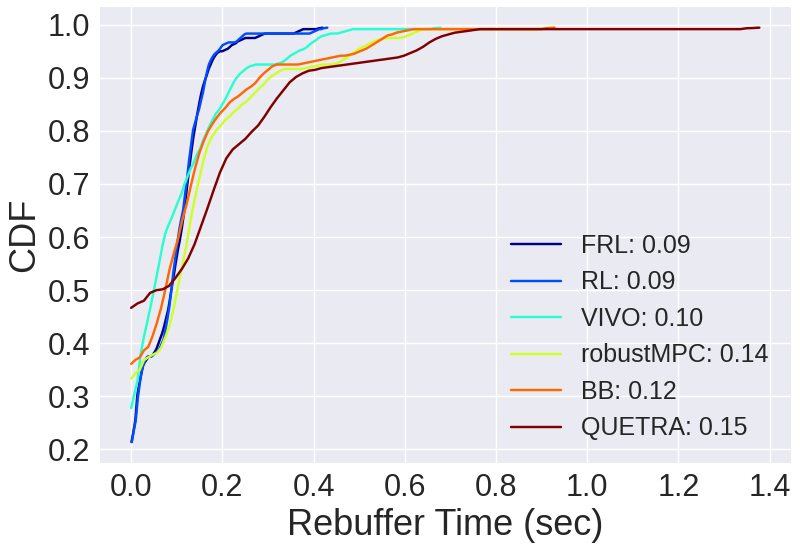}
       \caption{Rebuffering}
    \end{subfigure}
    \begin{subfigure}[h!]{0.32\textwidth}
       \includegraphics[width=\linewidth]{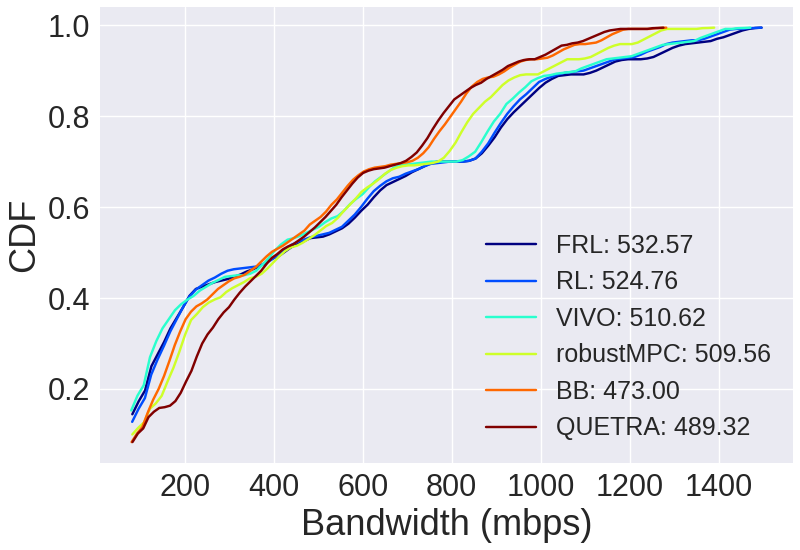}
       \caption{Bandwidth}
    \end{subfigure}
    \caption{Key metrics performance.}
    \label{fig:metrics}
\end{figure*}
\noindent \textbf{FOV prediction.}
Firstly, note that we only show the prediction results for the (X, Z) dimensions of the position and the (Yaw, Pitch) dimensions of the orientation, because users rarely move upward or downward (Y dimension), or rotate about the longitudinal axis (roll dimension), in the datasets we use as mentioned by ViVo~\cite{han2020vivo}. 
\autoref{fig:fov} shows the position MAE (\ref{fig:fovp}) and orientation MAE (\ref{fig:fovo}) of the three prediction methods described in \autoref{subsec:pre}. As shown, in every dimension, LR performs slightly better than MLP, and they both perform much better than LAST. Since LR has the best prediction accuracy while requires less computation than MLP, we choose to use LR in \sysname for FoV prediction.

\noindent \textbf{Network prediction.} 
As mentioned in \autoref{subsec:pre}, we tested two bandwidth prediction algorithms: GRU and EWMA. \autoref{fig:netprediction} shows their performances in two networking conditions, namely, while walking and driving. As shown, the EWMA-based method has a better performance. Besides, EWMA has a lower computational complexity than GRU. 
Therefore, we chose an EWMA-based method to predict the network bandwidth for better network prediction performance in \sysname. Specifically, we let the algorithm learn the bandwidth of the last 30~seconds to predict the next second.

\begin{figure*}[t!]
    \centering
       \includegraphics[width=.6\linewidth]{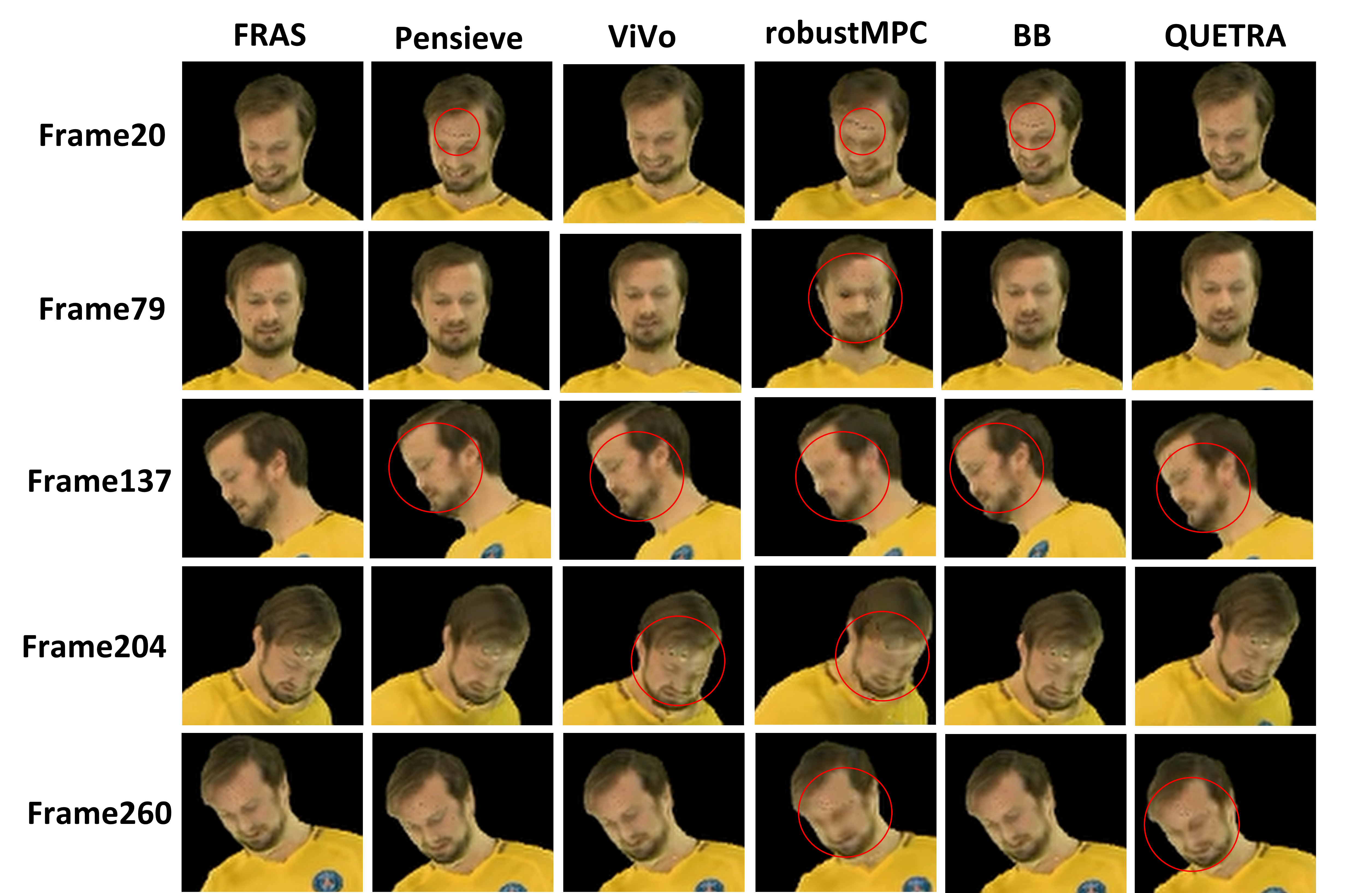}
    \caption{Demo results.}
    \label{fig:demo}
    \vspace{-0.1 in}
\end{figure*}

\vskip 0.05in \noindent \textbf{Training.} We first validate the core algorithm FRL by comparing its training performance with the RL algorithm adapted from Pensieve. \autoref{fig:training1} and \autoref{fig:training2} show the reward and loss of both algorithms during training. The curves of FRL represent the corresponding values of the global FL model while the curves of RL represent the corresponding values of an individual client model. The converged FRL outperforms Pensieve on reward by 9.77\% and loss by 49.81\%. 

As mentioned in Section \ref{sec:frl}, the number of local RL epochs, $E$, before each FL aggregation affects the performance tradeoff. We varied the data distribution heterogeneity across the clients and observed a pattern between the heterogeneity and optimal $E$ value. As shown in \autoref{fig:hetero}, as the heterogeneity increases, the algorithm requires a larger $E$ for good performance, which is reasonable since heterogeneous data distribution demands more local training to favor each individual dataset's pattern.

\vskip 0.05in \noindent \textbf{Test.} We validate the performance of FRL in comparison with the baselines using bandwidth traces collected in different networking conditions. As shown in \autoref{fig:testperform}, FRL significantly outperforms the baselines in all metrics under most scenarios. Specifically, FRL increases the average QoE (reward) by 86\%, 76\%, 68\%, 25\% and 22\% compared to QUETRA, robustMPC, BB, Pensieve, and ViVo, respectively. Moreover, FRL increases the average video quality level by 22\%, 16\%, 28\%, 5\% and 14\%, average PSNR by 5\%, 3\%, 6\%, 1\% and 4\%, bandwidth usage by 8\%, 4\%, 11\%, 1\% and 4\%, compared to QUETRA, robustMPC, BB, Pensieve and ViVo, respectively. Additionally, FRL decreases the average rebuffer time by 67\%, 56\%, 33\% and 11\% compared to QUETRA, robustMPC, BB and ViVo respectively. 

Note that we have overall 5 encoding quality levels (each level has a compressed and uncompressed version). Therefore, the improvements over baselines can be approximately seen as improving the average quality level, which aligns with the improvement results of PSNR. For the same reason, FRL presents higher bandwidth usage than the baselines. However, FRL presents the minimal rebuffering of all tested algorithms, which indicates FRL causes the lowest frequency of bandwidth saturation. Hence, it proves that FRL can best utilize the available bandwidth to achieve the best performance trade-off and QoE.

\vskip 0.05in \noindent \textbf{Demo.} We ran real-world tests for all the algorithms on the prototype system with point cloud videos on real wireless networks. \autoref{fig:demo} shows a sequence of point cloud video frames sampled approximately every two seconds. \sysname shows the best frame quality out of all the streaming algorithms, which validates the superiority of FRL and the practicability of \sysname.

\section{Conclusion}
Point cloud streaming plays a key role in multimedia, especially for volumetric videos. Its unique features, such as 6DoF and decoding complexity, demand innovative QoE definition and ABR algorithms. In this work, we propose \sysname, the first federated reinforcement learning framework, to the best of our knowledge, for adaptive point cloud video streaming. \sysname is an end-to-end framework that takes the unique features of the point cloud into account and augments learning performance by aggregating distributed modes of users but with privacy preservation. Empirical evaluations and prototype demo have shown the superior performance of \sysname in numerous perspectives. 
\label{sec:conclude}

\bibliographystyle{abbrv-doi}

\bibliography{ref}
\end{document}